\documentclass{article}
\usepackage[english]{babel}

\usepackage{lettrine}
\usepackage{amsmath}

\usepackage[colorlinks=true, allcolors=blue]{hyperref}
\usepackage[a4paper,left=20mm,right=20mm,top=15mm,bottom=20mm]{geometry} 
\usepackage{cite}
\usepackage{graphicx}
\usepackage{epstopdf}
\usepackage{fancyhdr}
\usepackage[T1]{fontenc}
\usepackage[utf8]{inputenc}
\usepackage{authblk}
\usepackage{multirow}
\usepackage{booktabs}

\newcommand{\figcaption}{\def\@captype{Fig. }} 
    {} 
\makeatother

\title{3D Orbital Angular Momentum Nonlinear Holography}

\author[1,$^\dagger$]{Feiyang Shen}
\author[2,$^\dagger$]{Weiwen Fan}
\author[2]{Yong Zhang}
\author[1,*]{Yuping Chen}
\author[1,3]{Xianfeng Chen}

\affil[1]{School of Physics and Astronomy,State Key Laboratory of Advanced Optical Communication Systems and Networks, Shanghai Jiao Tong University, 800 Dongchuan Road, Shanghai 200240, China}
\affil[2]{National Laboratory of Solid State Microstructures, College of Engineering and Applied Sciences, Nanjing University, Nanjing 210093, China}
\affil[3]{Collaborative Innovation Center of Light Manipulations and Applications, Shandong Normal University, Jinan 250358, China}

\affil[ ]{ }

\affil[ ]{Feiyang Shen:  shenfeiyang828@sjtu.edu.cn}
\affil[ ]{Weiwen Fan:  fww18021319944@163.com}
\affil[ ]{Yong Zhang:  zhangyong@nju.edu.cn}
\affil[ ]{Xianfeng Chen: xfchen@sjtu.edu.cn}
\affil[ ]{ }

\affil[ ]{Yuping Chen$^*$: ypchen@sjtu.edu.cn}
\affil[ ]{School of Physics and Astronomy, Shanghai Jiao Tong University,}
\affil[ ]{800 Dongchuan Road, Shanghai 200240, China}
\affil[ ]{Tel: +86-13816373910}
\affil[ ]{ }

\affil[$^\dagger$]{These authors contributed equally to this work.}

\date{} 

\begin{document}
\maketitle
\newpage
\begin{abstract}

Orbital angular momentum (OAM), due to its theoretically orthogonal and unbounded helical phase index, has been utilized as an independent physical degree of freedom for ultrahigh-capacity information encryption. However, the imaging distance of an OAM hologram is typically inflexible and determined by the focal length of the Fourier transform lens placed behind the hologram. Here, 3D orbital angular momentum holography is proposed and implemented. The Fourier transform between the holographic plane and imaging plane is performed by superimposing Fresnel zone plates (FZP) onto the computer-generated holograms (CGH). The CGH is binarized and fabricated on the birefringence lithium niobate crystal by femtosecond laser micromachining. Experimental verification demonstrates the feasibility of the encoding method. Moreover, by superimposing FZPs with different focal lengths into various OAM channels, OAM-multiplexing holograms are constructed. Target images are separately projected to different planes, thereby enabling 3D multi-plane holographic imaging with low crosstalk. The interval between adjacent imaging planes can be uniform and minimal, free from depth of field constraints, thus achieving high longitudinal resolution. This work achieves OAM holography in a more compact manner and further expands its applicability.

\end{abstract}

\section*{Introduction}
Computer-generated holograms(CGHs) utilizing precise amplitude and phase modulation provide an advanced method to design and manipulate complex light fields, including in 3D display\cite{dorsch1994fresnel,huang2013three,kang2019curved}, image projection\cite{meem2020multi}, beam shaping\cite{manousidaki20203d,zhu2020high}, and nonlinear holography\cite{zhu2020second,chen2023laser}, especially for Fresnel CGH, 
which can project multiple target images to arbitrary longitudinal position theoretically\cite{tu2023fabrication,makey2019breaking}. However, due to the varying depth of field (DoF) at the corresponding imaging plane, the interval between adjacent imaging planes is non-uniform, with larger intervals at more distant planes, leading to low longitudinal resolution and small capacity, or equivalently, strong crosstalk under the condition of dense image plane distribution (see Supplementary Note 1). Orbital angular momentum (OAM), as a new physical degree of freedom, has attracted intensive and diverse research interests\cite{fang2024orbital,wan2022toroidal}. It is promising in the multiplexing technique due to the orthogonality of its theoretically unbounded OAM modes\cite{fang2020orbital}. In traditional OAM holography, a cumbersome Fourier transform lens is typically placed behind the hologram, resulting in a fixed and inflexible imaging distance for an OAM hologram, thereby limiting its further application.
\par

\begin{figure}[h]
\centering\includegraphics[width=\linewidth]{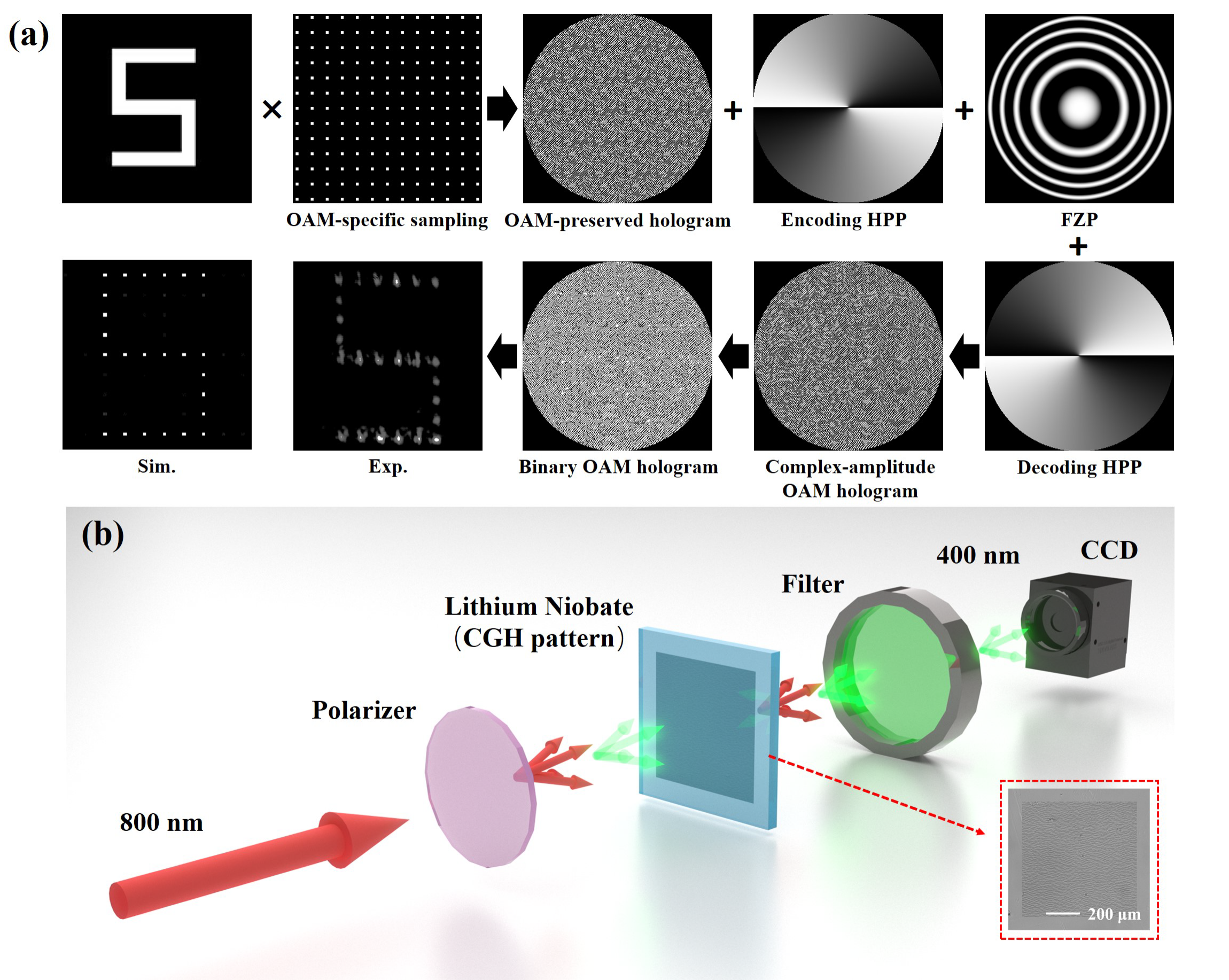}
\caption{(a) Design principle for the OAM hologram; (b) Schematic illustration of the optical system for holographic image reconstruction and the optical image of the fabricated hologram on LN sample by femtosecond laser.}
\end{figure}

In this paper, we propose and demonstrate a general approach to realize 3D orbital angular momentum nonlinear holography. By superimposing Fresnel zone plates onto CGHs, Fourier transform operations can be performed within the Fresnel regime\cite{makey2019breaking}, allowing flexible control of the projection depth in OAM holography by adjusting the focal length of the FZPs. Furthermore, the implementation of OAM-multiplexing hologram facilitates the independent control of the projection depth for each OAM channel, leading to the successful realization of low-crosstalk three-dimensional multi-plane imaging. Additionally, this holographic encoding technique obviates the necessity for the Fourier transform lens structure typically required in conventional OAM holography, thereby substantially simplifying the optical configuration. Femtosecond laser micromachining is becoming an advanced method used in various optical research\cite{wang2024laser,wang2024precise,li2020fib,liu2020nonlinear,zhu2023ultrafast}, offering an efficient and integrated fabrication approach.  The calculated CGH patterns are fabricated in monolithic lithium niobate crystals with binary modulation of quadratic susceptibility by femtosecond laser. Overall, our study contributes to the diversification of implementation methods for OAM holography. It also opens new possibilities for nonlinear multi-plane holography in a single piece of crystal.
  
\section*{Result}
\subsection*{Design Principle of 3D OAM Nonlinear Holography}
The encoding process of the OAM holography is shown in Fig. 1a. The target image is firstly sampled according to the spatial frequency of the selected encoding helical phase (see Supplementary Note 2). Then, the Gerchberg-Saxton (GS) algorithm is utilized to obtain the OAM-preserved hologram. To construct an OAM-selected hologram, a helical phase plate (HPP) with $\textit{l}_c$ is encoded onto the OAM-preserved hologram. According to the nonlinear OAM conservation law\cite{hu2020nonlinear} and nonlinear OAM-matching condition\cite{fang2020multichannel} $\Delta \textit{l}=2\textit{l}_w+\textit{l}_c$, where $\textit{l}_w$ is the topological charge carried by the incident fundamental beam. The target image can be clearly reconstructed only when an incident OAM beam with $\textit{l}_w=-1/2\textit{l}_c$ is used and the reconstructed image consists of pixels that feature Gaussian mode showing high-intensity distribution. On the contrary, if a mismatched OAM beam is incident, each pixel maintains the vortex characteristics and exhibits a lower intensity distribution, which can be regarded as background noise. In this work, a decoding HPP with $-\textit{l}_c$ is directly superimposed onto the hologram instead of the OAM beam incidence.
\par

\begin{figure}[h]
\centering\includegraphics[width=\linewidth]{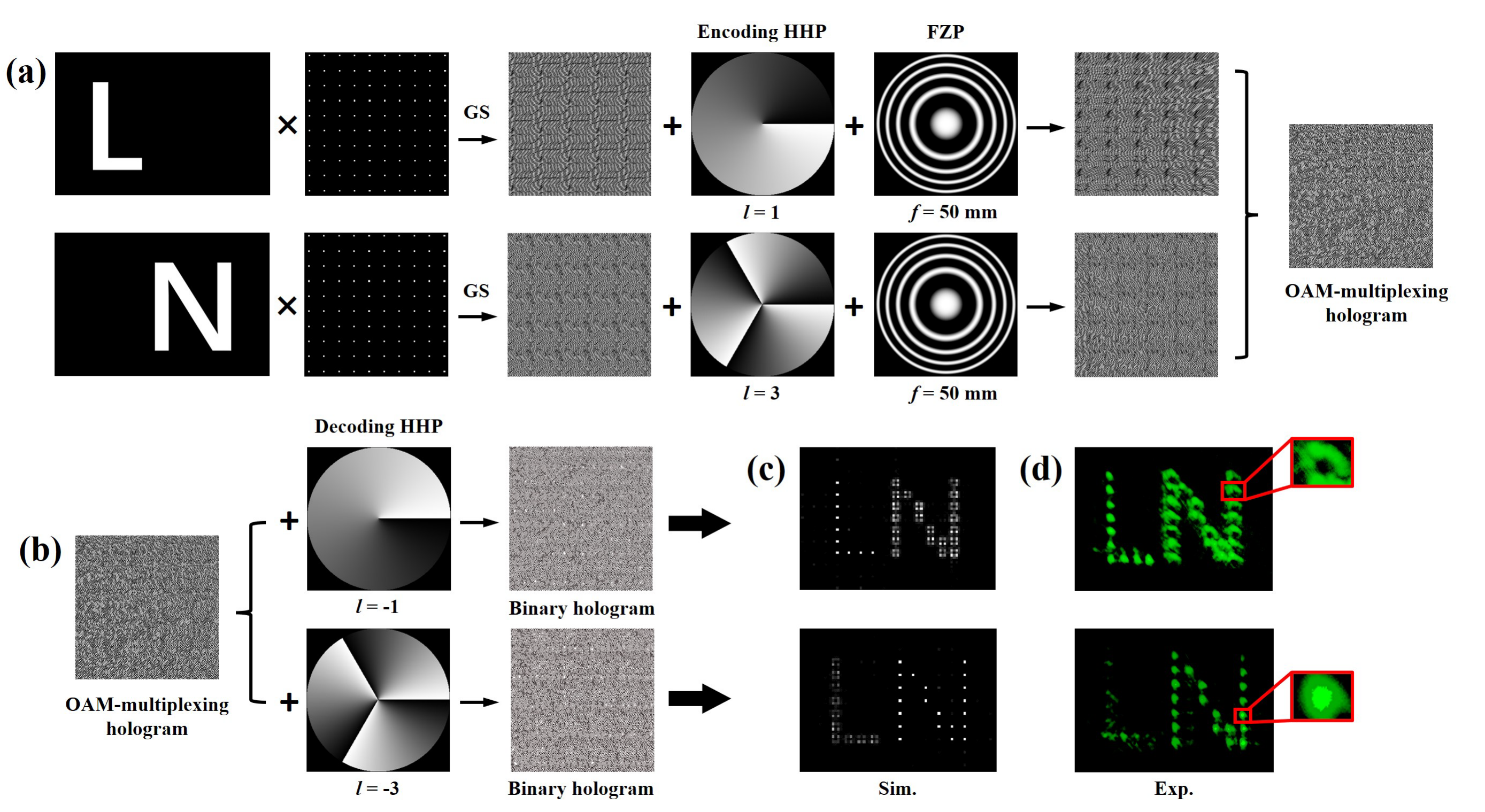}
\caption{(a) Design principle for the OAM-multiplexing hologram with the same depth; (b) The decoding process by superimposing corresponding decoding HHP; (c) Simulated and (d) Experimental reconstruction results at $z=50 mm$.}
\end{figure}

In the OAM holography, the electric field on the hologram plane and that on the image plane form a Fourier pair. Direct superposition of a FZP on a Fourier hologram will generate a single-plane Fresnel hologram (see Supplementary Note 3), where the focal length of the FZP can be used to translate the image to any distance\cite{makey2019breaking} controllably. Notably, this flexibility is further exemplified by our ability to superimpose Fresnel zone plates with different focal lengths in various OAM channels, thereby projecting different target images onto planes at distinct depths, namely, the three-dimensional multi-plane OAM holography. Meanwhile, the traditional Fourier transform lens structure can be omitted, which greatly simplifies the optical configuration. The phase function of FZP can be written as:
\begin{equation}
    \Phi_{FZP}(x,y)=-\frac{k}{2f}(x^2+y^2)
\end{equation}
where $(x, y)$ represents the Cartesian coordinates in the holographic plane, $k$ is the wavevector and $f$ represents the focal length. So the final complex-amplitude distribution of an OAM multiplexed hologram can be described mathematically as:
\begin{equation}
    U(x,y)=exp(i\Phi_{dec}(x,y))\sum_{j=1}^{n}E_j(x,y)exp(i\Phi_{j-FZP}(x,y))exp(i\Phi_{j-enc}(x,y))
\end{equation}
where n is the total OAM channel number, $E_j(x,y)$ is the complex-amplitude distribution of the sampled target image obtained by GS algorithm encoded on the j-th OAM channel, $\Phi_{j-enc}(x,y)$ is the phase distribution of the j-th encoding HPP, $\Phi_{j-FZP}$ denotes the phase function of the FZP superimposed on the j-th OAM channel and $\Phi_{dec}(x,y)$ is the phase distribution of the decoding HPP. It is worth noting that the depth controllability is achieved by loading different FZP phases onto different OAM channels. It should be emphasized that the size of the reconstructed target image is directly related to the focal length of the FZP. The larger the focal length, the larger the size of the image. Therefore, to achieve a realistic restoration of a three-dimensional object, the images of each plane need to be scaled accordingly in advance (Supplementary Note 4). 
\par
Finally, we choose a suitable way to encode the complex-amplitude field onto the holographic plane. Considering that we use the femtosecond laser to erase the nonlinear coefficient of the lithium niobate crystal which results in two states corresponding to the actual laser-nonirradiated area and laser-irradiated area, the binary Fresnel CGH was selected in our experiments (see Methods). 

\begin{figure}[h]
\centering\includegraphics[width=\linewidth]{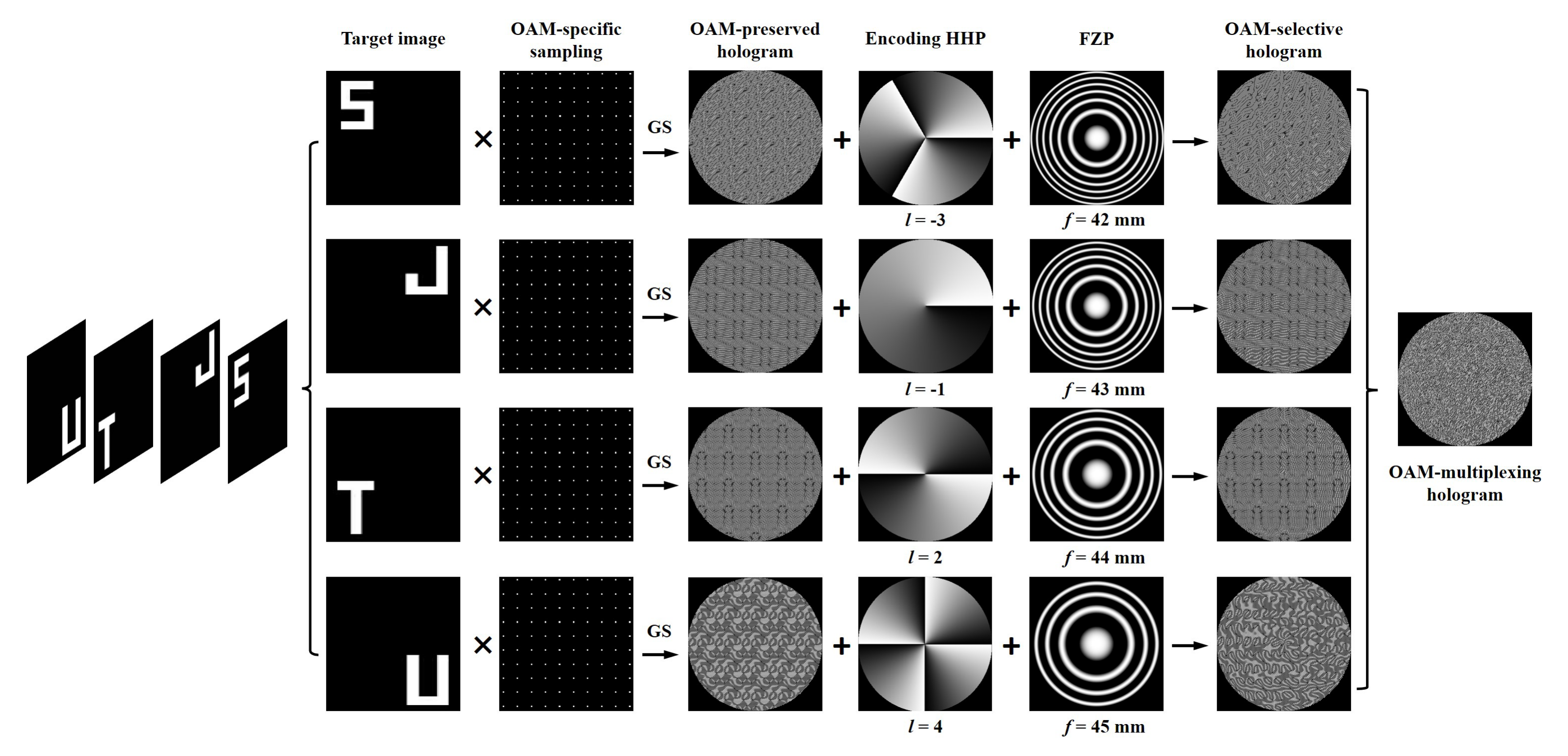}
\caption{Encoding process of OAM-multiplexing holography for 3D multi-plane imaging. The three-dimensional object "SJTU" is sliced into four planes containing different letters. Each letter is encoded into a distinct OAM channel by applying OAM-specific sampling before performing the GS algorithm and then superimposing with the phase of encoding HHP and FZP. Consequently, four OAM-selective holograms are obtained respectively and then superimposed to construct an OAM-multiplexing hologram.}
\end{figure}

Notably, both the FZPs and the binarization process in the encoding process are wavelength-dependent. Therefore, we adopted a configuration where the beam was first frequency-doubled and then modulated, with the design of CGHs specifically tailored for the SH beam. Under the irradiation from the opposite direction, the target image can not be clearly reconstructed(see Supplementary Note 5), enabling nonreciprocal information processing.
\par
To validate the effectiveness of this encoding method, the letter S was encoded with a HPP with $l=4$ and a FZP with $f=50 mm$. When a decoding HPP with $l=-4$ was superimposed onto the hologram, a distinct SH pattern emerged at the corresponding distance with Gaussian-spot pixels under the condition of Gaussian beam incidence, as shown in Figure 1a. The slight difference between the experiment and theoretical simulation is the imperfection of the produced ferroelectric domain structure and the unavoidable random structure errors.

\begin{figure}[h]
\centering\includegraphics[width=\linewidth]{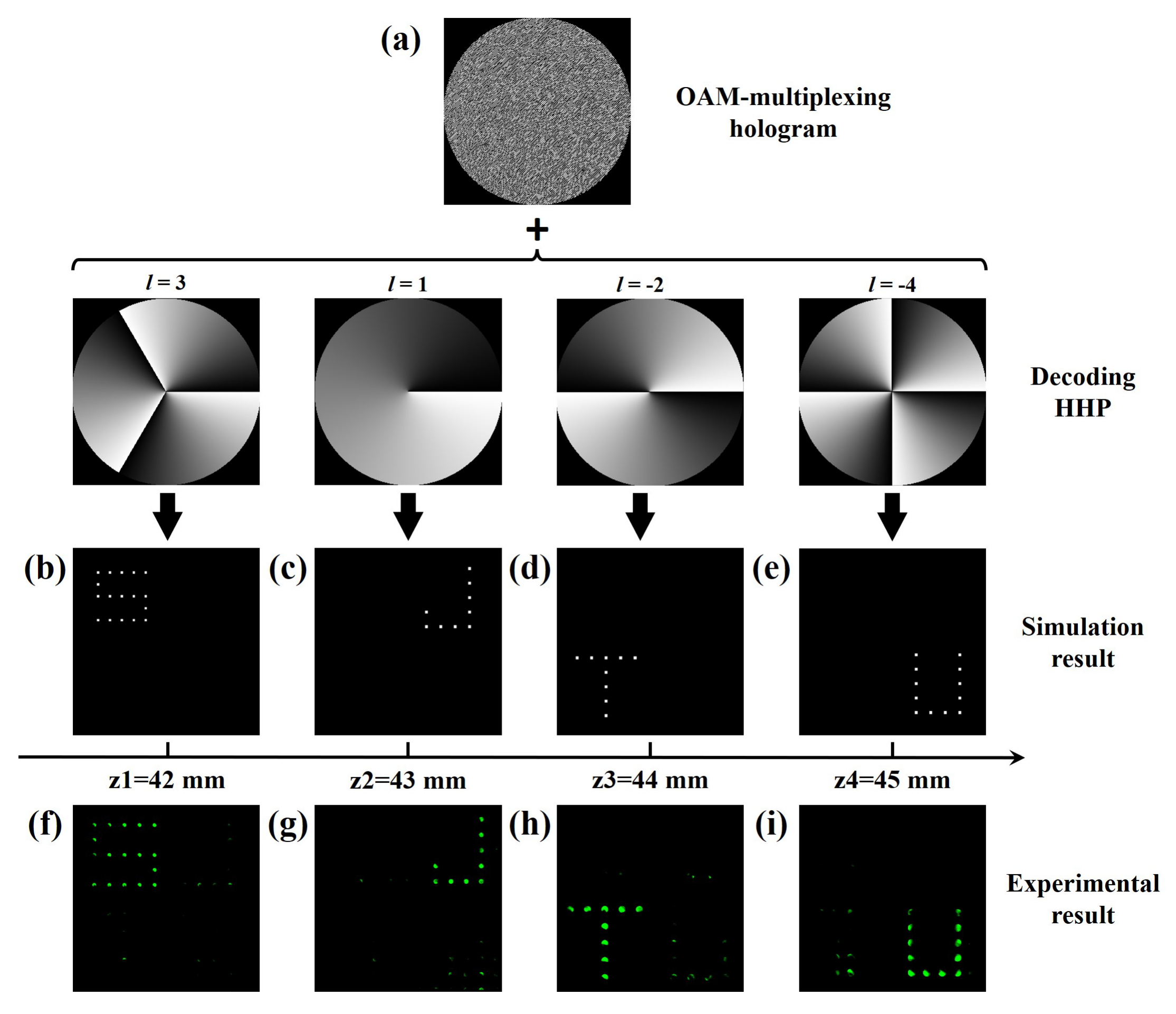}
\caption{(a)Decoding process of OAM-multiplexing holography for 3D multi-plane imaging; (b)-(e) Numerical reconstructions of 3D model “SJTU” at the depth of z1, z2, z3 and z4, respectively; (f)-(i) Experimentally reconstructed images at corresponding positions.}
\end{figure}

\subsection*{OAM-multiplexing holography at the same imaging plane}
The more pronounced difference between the OAM-matching and OAM-mismatching can be demonstrated by constructing an OAM-multiplexing hologram. As can be seen from Figure 2a, letters L and N, namely the abbreviation for lithium niobate crystal, were encoded separately into two OAM channels, i.e. $l = 1$ and $l = 3$, while superimposing identical FZPs with $f = 50 mm$ projects them onto the same imaging plane. The decoding process can be performed by superimposing corresponding decoding HPP onto the OAM-multiplexing hologram as shown in Figure 2b. Figure 2c and (d) show the simulation and experimental results, respectively. It is clear that the decoded target image consisted of Gaussian-shaped pixels showing high intensity, corresponding to the OAM-matching condition being satisfied. Due to the OAM mismatching, the other letter had donut-shaped pixels and relatively low intensity, which can be regarded as background noise and can be further removed through post-processing.
\subsection*{OAM-multiplexing holography for 3D multi-plane imaging}
\par

Suffering from strong crosstalk and varying DoFs between adjacent planes, it is challenging for traditional 3D holographic imaging techniques to achieve large-capacity and high-fidelity 3D information storage. By leveraging the orthogonality of OAM modes, crosstalk can be significantly reduced, especially since the proposed encryption method now enables precise and independent control over the imaging distances of each OAM channel. Additionally, the intervals between adjacent imaging planes are no longer constrained by the DoFs. 

\par

To characterize the 3D reconstruction performance of the proposed OAM holography, we chose multi-plane images of the letters "SJTU" as our 3D object model. As is illustrated in Figure 3, the four letters are divided into four layers which are positioned at different distances from the hologram ($z_1=42mm$, $z_2=43mm$, $z_3=44mm$, $z_4=45mm$). After performing the GS algorithm, they were separately encoded into four different OAM channels ($l_1=-3$, $l_2=-1$, $l_3=2$, $l_4=4$) and superimposed with corresponding FZPs for independent projection. The OAM-multiplexing hologram was obtained by superposing four DC-OAM-selective holograms. The decoding process is shown in Figure 4a. To reconstruct a specific target image, correct decoding HHP should be superimposed onto the OAM-multiplexing hologram. The numerical reconstructions under the Gaussian beam incidence at four different depth planes are shown in Figure 4b-e and the corresponding experimental reconstruction results are shown in Figure 4f-i. After applying a mode-selective aperture array in post-processing of the reconstructed holographic images, most of the OAM-mismatching target images were filtered. As we can see, the expected letters appeared at pre-designed distances. Due to inevitable discrepancies between the actual fabricated structures and the designed CGH pattern, the interference between OAM modes in adjacent pixels, termed as adjacent pixel interference (API), is unavoidable. Consequently, high-intensity pixels comparable to those in the signal regions appear in the non-signal areas during experiments, and these pixels cannot be filtered out.

\section*{Discussion}
The imaging distances and the number of target images selected in this experiment are merely exemplary. In practice, smaller distance intervals and a greater number of images can be selected. Both of the transverse resolution and longitudinal resolution, or equivalently the capacity, are constrained by the OAM sampling condition and the numerical aperture of the CCD camera. A larger number of target images necessitates a larger sampling constant. This implies that the target images need to be magnified and a camera with a larger numerical aperture should be used; otherwise, the resolution of the images will degrade, potentially leading to distortion. If our holograms are loaded onto a spatial light modulator and a temporal multiplexing method is employed\cite{shi2023super}, this issue can be mitigated. A more detailed discussion of the capacity and resolution is provided in Supplementary Note 6. Meanwhile, our encryption method is compatible with other enhanced OAM holography techniques\cite{cheng2021ellipticity,zhu2021ultra,wang2022enhancing,wang2022angular,li2022multiple,li2023multiple,zhang2023multiparameter,shi2023super}, and their integration allows us to further enhance the capacity and resolution of three-dimensional holographic imaging.

In conclusion, we have proposed the concept of 3D OAM nonlinear holography and employed femtosecond laser micromachining technique to fabricate CGH pattern onto the lithium niobate crystal. Our approach utilizes superimposed FZPs in CGH replacing the Fourier transform lens used in the traditional OAM holography, enabling us to flexibly control over imaging distance. Meanwhile, the optical configuration is greatly simplified. Moreover, we constructed OAM-multiplexing holograms to independently project images that loaded in different OAM channels to distinct positions, thereby achieving low-crosstalk 3D multi-plane holographic imaging. Experimental verification has demonstrated the feasibility of 3D OAM nonlinear holography. It is worth mentioning that the proposed OAM holographic encoding strategy represents a universal approach compatible with the majority of OAM holography techniques, holding great promise for applications in various field including 3D display and optical manipulation.

\section*{Method}
\subsection*{Binarization of the Complex-amplitude OAM Hologram}
To record the information of the complex-amplitude OAM hologram onto the lithium niobate crystal, the method of binary interference CGH is used. Through the interference of the object wave $U(x,y)=A_0(x,y)exp(i\phi_0(x,y)$ and the planar reference wave $R(x, y) = R \exp(i 2 \pi \alpha x)$, where $A_0(x,y)$ and $exp(i\phi_0(x,y))$ are the amplitude and phase term of the complex amplitude on the holographic plane, $R$ and $\alpha$ are the amplitude and the carrier frequency respectively, we first obtain the transmittance function $h(x,y)$ of the hologram:

\begin{equation}
    \begin{split}
        h(x,y) &= {\mid U(x,y) + R(x,y) \mid}^2 \\
               &= \frac{1}{2} \{ 1 + A_0(x,y) [ \cos \phi_0(x,y) - 2 \pi \alpha x ] \}
    \end{split}
\end{equation}
The bright fringes equation of the hologram can be obtained as:
\begin{equation}   
        \phi_0(x,y)-2 \pi \alpha x=2 \pi n
\end{equation}
where \( n \) represents the serial number of the bright fringes, \( n = 0, \pm 1, \pm 2, \ldots \).By solving the position of each bright fringe and opening a thin slit, a binary transmission grating is formed, that is, the phase of the object wave is encoded. The encoding of the amplitude is obtained by introducing an offset $cos\pi q(x,y)$ to modulate the width of the bright fringes\cite{lee1974binary}, where $q(x,y)=arcsin[Ao(x,y)]/\pi$ and the amplitude $Ao(x,y)$ takes the normalized value. This encoding method is widely used in standard wavefront generation and interference detection\cite{wyant1972using,burge1995applications}. Then, the final interference binary CGH is represented as follows:
\begin{equation} 
H(x, y) =
\begin{cases} 
      1, & \cos[\phi_0(x, y) - 2\pi \alpha x] \geq \cos \pi q(x, y) \\
      0, & \text{others}
\end{cases}
\end{equation}

\subsection*{Fabrication of CGH by Femtosecond Laser Erasure}
The CGH is fabricated using a NIR laser working at $800 nm$ wavelength, $1 kHz$ reputation rate, and $109 fs$ pulse width (Coherent Legend Elite). A $5 mol.\%$ MgO-doped LiNbO3 crystal with a thickness of $1 mm$ is used as a sample which is mounted on the computer-controlled XYZ translation stage with a resolution of $0.2 \mu m$. An objective lens with a numerical aperture of $0.90$ (Nikon CFI TU Plan Fluor BD) is applied to focus the laser pulse onto the surface of the crystal. The moving speed is $100 \mu m/s$ and the moving direction is perpendicular to the laser beam. In our experiment, a single pulse of about $30 \mu J$ is used. Controlled by the computer program, the ferroelectric domain of the crystal is selectively erased corresponding to the dark area of the CGHs. While the non-irradiated points correspond to the white area of the CGHs. The physical mechanism of the ferroelectric domain erasing process can be understood in that the crystallinity is reduced through laser irradiation\cite{wei2018experimental}.We fabricated the CGH patterns with $300\times300$ pixels, where each pixel has a size approximately at $2 \mu m  \times  2 \mu m$. The fabricated CGHs were written within an area of $0.6 \times 0.6 mm^2$ and the total processing time is $1.5h$.
\subsection*{Experimental Set-up
}
Figure 1b shows the optical holographic imaging system. The fundamental frequency (FF) beam with a wavelength of $800 nm$ is used for image reconstruction. A linear polarizer was utilized to adjust the polarization of the FF beam. Afterward, an ordinary polarized laser beam was irradiated onto the sample. Given that the CGH pattern was inscribed on the surface of the LiNbO3 crystal, we arranged the sample such that the FF beam initially traversed the unmodulated LiNbO3 crystal, where the second harmonic (SH) beam was generated at the wavelength of $400 nm$. Subsequently, the SH beam was modulated by the CGH and projected to the corresponding depth for imaging. Before the CCD camera, a filter was used to separate the FF and SH beams. 
\section*{Acknowledgement}
This work was supported by the National Natural Science Foundation of China (Grant Nos. 12134009), and Shanghai Jiao Tong University (SJTU) (Grant No. 21X010200828).

\section*{Author contributions}

Feiyang Shen prepared the manuscript in discussion with all authors. Feiyang Shen and Yuping Chen designed the holographic method and analyzed the data. Feiyang Shen and Weiwen Fan fabricated the hologram onto the lithium niobate crystal and performed the experiments. Feiyang Shen, Yuping Chen and Yong Zhang revised the manuscript. Yuping Chen supervised the project.

\section*{Conflict of interest}

The authors declare no competing interests.

\bibliography{sample}
\bibliographystyle{unsrt}

\end{document}